\documentclass[10pt]{iopart}
\usepackage{epsf}  
\begin{document}

\title[Dilepton production in proton-proton collisions at BEVALAC energies]
{Dilepton production in proton-proton collisions at BEVALAC energies}

\author{Amand Faessler\dag, 
Christian Fuchs\ddag
\footnote[3]{E-mail address: christian.fuchs@uni-tuebingen.de},\\
Mikhail Krivoruchenko\dag\ddag $\;$and Boris Martemyanov\dag\ddag}

\address{\dag\ Institut f\"{u}r Theoretische Physik, Universit\"{a}t T\"{u}bingen\\
Auf der Morgenstelle 14, D-72076 T\"{u}bingen, Germany}

\address{\ddag\ Institute for Theoretical and Experimental Physics \\
B. Cheremushkinskaya 25, 117259 Moscow, Russia\medskip}

\begin{abstract}
The dilepton production in elementary ${pp\rightarrow e^{+}e^{-}X}$
reactions at BEVALAC energies $T_{lab}=1\div 5$ GeV is investigated. The
calculations include direct ${e^{+}e^{-}}$ decays of the vector mesons $\rho
^{0}$, $\omega $, and $\phi $, Dalitz decays of the $\pi ^{0}$-, $\eta $-, $%
\rho $-, $\omega $-, and $\phi $-mesons, and of the baryon resonances $%
\Delta (1232),N(1520),$ $\dots $ . The subthreshold vector meson production
cross sections in $pp$ collisions are treated in a way sufficient to avoid
double counting with the inclusive vector meson production. The vector meson
dominance model for the transition form factors of the resonance Dalitz
decays $R\rightarrow e^{+}e^{-}N$ is used in an extended form to ensure correct asymptotics
which are in agreement with the quark counting rules.  Such a modification gives an
unified and consistent description of both $R\rightarrow N\gamma $
radiative decays and $R\rightarrow N\rho (\omega )$ meson decays.
 The effect of
multiple pion production on the experimental efficiency for the detection of
the dilepton pairs is studied. We find the dilepton yield in reasonable
agreement with the experimental data for the set of intermediate energies 
whereas at the highest energy $T_{lab}=4.88$ GeV the number of 
dilepton pairs is likely to be overestimated experimentally 
in the mass range $M=300\div 700$ MeV. 

\end{abstract}

\pacs{25.75.Dw, 13.30.Ce, 12.40.Yx}

\maketitle

\section{Introduction}

Relativistic heavy-ion collisions present an unique possibility to create
nuclear matter at high densities and temperatures where the hadron
properties become different and a phase transition to quark matter with
signatures of the deconfinement and restoration of chiral symmetry is
expected. The change of the nucleon mass in the nuclear matter was
implemented into the Walecka model \cite{Wal,Chin} already in 1970's in the
framework of the effective hadron field theory. Later on this effect was put
on the firmer grounds on the basis of a partial restoration of the chiral
symmetry and finite-density QCD sum rules \cite{DRUC,MES}, the
Nambu-Jona-Lasinio model \cite{MAR}, and effective meson field theory 
\cite{Chin,BR,brown96}.

Dileptons ($e^{+}e^{-},$ $\mu ^{+}\mu ^{-}$) are the most clear probes of
the high density nuclear matter. The reason is that the dileptons interact
with the matter only by electromagnetic forces and can therefore leave the
heavy-ion reaction zone essentially undistorted by final state interactions.
They provide valuable information on the in-medium properties of
hadrons and, hence, on the state of matter.

The dilepton spectra from heavy-ion collisions have been measured
at two different energy scales: 
by the CERES and HELIOS-3 Collaborations at SPS 
\cite{CERES,HELIOS} (a few hundreds GeV per nucleon) and by the DLS
Collaboration at BEVALAC \cite{BEVALAC} (a few GeV per nucleon). In the
CERES and HELIOS-3 experiments and in the BEVALAC experiment the production
of dileptons with invariant masses between $300\div 700$ MeV is found to be 
enhanced as compared to estimates based on the theoretically known dilepton
sources when in-medium modifications of hadron properties are neglected.

The data on the total photoabsorption cross section on heavy nuclei 
\cite{fras} give an evidence for the broadening of nucleon resonances 
in the nuclear medium \cite{Kon}. The physics behind is the same as in the
collision broadening of the atomic spectral lines in hot and dense gases,
discussed by Weisskopf in the early 1930's \cite{Wei} 
(for the present status of
this field, see, e.g., \cite{SOB}). This rather general effect which 
takes its origin from the atomic spectroscopy should also lead to a broadening 
of the $\rho $-mesons in heavy-ion collisions. 

The low-energy dilepton excess can be explained by a reduction of the 
$\rho$-meson mass in a dense medium \cite{Koch,Li,WC}. The in-medium modification 
of the $\rho$-meson spectral function \cite{RAPP,PPLLM} leading to a 
broadening of the $\rho$-meson seems also to be sufficient to account for 
the CERES and HELIOS-3 data \cite{WC}. 

In the DLS experiment a different temperature and density regime is probed.
The enhancement of the dilepton spectra due to the reduction of the $\rho $%
-meson mass and the $\rho $-meson broadening is not sufficient to bring 
theoretical estimates in coincidence with the available 
experimental data \cite{BK}. The
in-medium $\rho $-meson scenarios that successfully 
explain the dilepton yield at SPS
energies fail for the DLS data. This phenomenon was called the ''DLS
puzzle''. It worthwhile to notice that the final data from the DLS
Collaboration have changed by about a factor of $5-7$ as compared to the
initially reported results. The future HADES experiment at GSI will study
the dilepton spectra at the same energy range in greater details \cite{friese}.

A possibility to clarify the origin of the DLS puzzle has appeared since
data from elementary $pp$ $(pd)$ collisions at $%
T=1\div 5$ GeV ($T$ is the kinetic energy of the incident proton in the
laboratory frame) became available from the DLS Collaboration \cite{W}. The
elementary cross sections enter as an input into the transport 
simulations of heavy-ion collisions, so their better understanding is of
great value.

The dilepton spectra in the $pp$ collisions at $T=1\div 5$ GeV have been
calculated in refs.\cite{ernst,BCEM,BCM}. In ref.\cite{ernst} the agreement
achieved with the DLS\ data is generally good at low energies where the
subthreshold production of nucleon resonances is
important. When the energy increases and the inclusive production becomes
dominant, the dilepton yield is underestimated at the same mass range $%
300\div 700$ MeV as in the heavy-ion collisions. A signature for this
effect exists also in the calculations of refs.\cite{BCEM,BCM}. 
This can be interpreted to mean that the studies \cite{ernst,BCEM,BCM} 
revealed, apparently, the reoccurrence of the DLS puzzle on 
the elementary level of
the nucleon collisions. They leave, therefore, a doubt on the quality of the
experimental data and/or the reliability of the accepted theoretical schemes.

This paper is devoted to a further going theoretical analysis 
of the elementary 
dilepton production cross sections.

In the next Sect., the production mechanisms are critically revisited. 
The subthreshold production 
cross sections for the vector mesons are treated such that no double counting
appears with the inclusive processes. The effect of 
multiple pion production on the experimental detection efficiency 
for the dilepton pairs is also studied. We demonstrate 
that the detector efficiency is
sensitive to the number of pions produced and propose a simple model to
account for the multiple pion production effects.

In the Dalitz decays of the nucleon resonances, ${R\rightarrow Ne^{+}e^{-}}$%
, the Vector Meson Dominance (VMD) model is usually applied for the description of
the resonance transition form factors. However, the naive VMD which takes the $%
R\rightarrow N\rho $ data as an input systematically overestimates the radiative decay
branchings. In Sect.3, the VMD model is extended to ensure the correct asymptotic
behavior of the transition form factors in agreement with the quark counting
rules. Such a modification is found to be sufficient to achieve an 
unified description of both, $R\rightarrow
N\gamma $ radiative decays and $R\rightarrow N\rho (\omega )$ meson decays. 
Our estimates of the subthreshold cross sections
rely therefore on the two essentially different sets of the experimental
data, $R\rightarrow N\gamma $ and $R\rightarrow N\rho $.
The numerical results are discussed in Sect.4. We found that the above
improvements do not eliminate the discrepancy with the DLS\ data at $T=4.88$
GeV. Moreover, the results for the lowest energy, $T=1.04$ GeV, also require
an additional study from the experimental and/or theoretical side.

\section{${pp\rightarrow }e^{+}e^{-}{X}$ reaction}

The dilepton production in nucleon collisions goes through the production of
virtual photons which decay subsequently into $e^{+}e^{-}$ pairs. According
to the VMD model, the virtual photons are coupled
to vector mesons $V=\rho ^{0},$ $\omega,$ and $\phi .$ The dilepton
production can therefore be calculated using the inclusive vector meson
production cross sections: 
\begin{equation}
\frac{d\sigma (s,M)^{pp\rightarrow e^{+}e^{-}X}}{dM^{2}}=\sum_{V}(1+n_{V})%
\frac{d\sigma (s,M)^{pp\rightarrow VX}}{dM^{2}}B(M)^{V\rightarrow e^{+}e}.
\label{V1}
\end{equation}
Here, $s$ is the square of the invariant mass of two colliding protons, $M$
the invariant mass of the dilepton pair, $d\sigma (s,M)^{pp\rightarrow
VX}/dM^{2}$ is the differential  vector meson production cross
section, and $n_{V}$ is the average number of additional vector mesons $V$
in the state $X$. In the energy range of interest, $T=1\div 5$ GeV,
where $T$ is the kinetic energy of the proton in the laboratory
system, $n_{V}=0$. The branching ratio 
\begin{equation}
B(M)^{V\rightarrow e^{+}e}=\frac{\Gamma (M)^{V\rightarrow e^{+}e}}{\Gamma
_{tot}^{V}(M)}  \label{BR}
\end{equation}
corresponds to the direct $V\rightarrow e^{+}e^{-}$ decays, with $\Gamma
_{tot}^{V}(M)$ being the total meson decay width.

In order to disentangle the various contributions, we decompose the cross section 
entering into eq.(\ref{V1}) into pole and background parts: 
\begin{equation}
d\sigma (s,M)^{VX}=d\sigma (s,M)_{P}^{VX}+d\sigma (s,M)_{B}^{VX}.  \label{V2}
\end{equation}
Such a decomposition implies that interference effects between
the different sources are neglected. The background sources like $\pi \rightarrow
\gamma e^+e^-$ and $\eta \rightarrow \gamma e^+e^-$ do not interfere due to 
the kinematical reasons. These two reactions in turn do not interfere with 
the reaction $\omega \rightarrow \pi^0 e^+e^-$, since the final states are 
different. The reactions $NN \rightarrow RN$, $R \rightarrow e^+e^-X$
going through the nucleon resonances $R$ with different quantum numbers
do not interfere with each other either. The sources like 
$NN \rightarrow N\Delta(1232)$, $\Delta(1232) \rightarrow Ne^+e^-$ 
and $NN \rightarrow NN\rho$, $\rho \rightarrow e^+e^-$ do interfere, however.
The relative phases of the amplitudes describing the different reactions are 
unknown, so the neglection of the interference between all the 
reactions constitutes a reasonable first approximation. 

The distribution over the meson mass $M$ in the pole part of the cross
section has a Breit-Wigner form corrected to the available phase space for
the final state $VX$. At moderate energies, the state $X$ is dominated by
two nucleons and pions, so one can write 
\begin{eqnarray}
d\sigma (s,M)_{P}^{VX}&=&\sigma (s)_{P}^{VX}\frac{1}{\pi }\frac{M\Gamma
_{tot}^{V}(M)dM^{2}}{(M^{2}-m_{V}^{2})^{2}+(M\Gamma _{tot}^{V}(M))^{2}} \nonumber \\
&\times&\sum_{n=0}^{N}w_{n}C_{n}\Phi _{3+n}(\sqrt{s}...)  \label{V3}
\end{eqnarray}
where 
\begin{equation}
\Phi _{3+n}(\sqrt{s}...)=\Phi (\sqrt{s},m_{N},m_{N},M,\mu _{\pi },...,\mu
_{\pi })  \label{phase_space}
\end{equation}
is the $(3+n)$-body phase space of the final state (two nucleons with masses 
$m_{N}$, one vector meson $V$ with mass $m_{V},$ and $n$ pions with masses $%
\mu _{\pi }$). The value $N_{\pi }=\left[ (\sqrt{s}-2m_{N}-m_{V})/\mu _{\pi
}\right] $ is the maximal number of pions allowed by energy conservation, $%
\left[ x\right] $ denotes the integer value of $x$, and the values $w_{n}$
are the probabilities for the production of $n$ pions, with 
\begin{equation}
\sum_{n=0}^{N_{\pi }}w_{n}=1.  \label{sw=1}
\end{equation}
The normalization factor $C_{n}$ is given by 
\begin{equation}
C_{n}^{-1}=\int_{\mu _{0}^{2}}^{(\sqrt{s}-2m_{N}-n\mu _{\pi })^{2}}\frac{1}{%
\pi }\frac{M\Gamma _{tot}^{V}(M)dM^{2}}{(M^{2}-m_{V}^{2})^{2}+(M\Gamma
_{tot}^{V}(M))^{2}}\Phi _{3+n}(\sqrt{s}...)  \label{V4}
\end{equation}
where $\mu _{0}$ is the physical threshold for vector meson decays ($\mu
_{0}=2\mu _{\pi }$ for the $\rho $-meson). Notice that the cross section (%
\ref{V2}) vanishes at values $M<$ $\mu _{0}$. However, the total width $%
\Gamma _{tot}^{V}(M)$ entering into the denominator of the branching ratio (%
\ref{BR}) at $M<$ $\mu _{0}$ vanishes as well, so that the cross section ( 
\ref{V1}) is actually finite everywhere above the two-electron threshold.

In the zero-width limit, $\Gamma _{tot}^{V}(M)=0,$ eq.(\ref{V3}) simplifies
to give 
\begin{equation}
d\sigma (s,M)_{P}^{VX}=\sigma (s)_{P}^{VX}\delta (M^{2}-m_{V}^{2})dM^{2}.
\label{zero-width}
\end{equation}
The finite-width effects are important for the $\rho $-meson and less
important for $\omega $- and $\phi $-mesons.

As the background we consider states with $\pi $-mesons
originating from the $V$-meson strong decays, that do not contribute to
the pole part of the cross section of  $V$-meson production. We assume {\it e.g.}
that the $\rho $-mesons from the reactions 
$$
\pi ^0\rightarrow \gamma \rho ^0\rightarrow \gamma e^{+}e^{-}\;(\gamma \pi
^{+}\pi ^{-}), 
$$
$$
\eta \rightarrow \gamma \rho ^0\rightarrow \gamma e^{+}e^{-}\;(\gamma \pi
^{+}\pi ^{-}), 
$$
$$
\omega \rightarrow \pi ^0\rho ^0\rightarrow \pi ^0e^{+}e^{-}\;(\pi ^0\pi
^{+}\pi ^{-}), 
$$
$$
R\rightarrow N\rho ^0\rightarrow Ne^{+}e^{-}\;(N\pi ^{+}\pi ^{-}) 
$$
form a background, which should be added to the pole part of 
$\rho$-mesons production. (In the last line only the $\rho$-mesons 
away from the resonance region correspond to background.) 

The dilepton invariant mass in all these reactions can be lower than the two-pion 
threshold. Clearly, all processes with $M<2\mu _\pi $ are not accounted for by 
the pole part of the inclusive cross section. Their contributions should be added 
to the contribution of the pole part.
With increasing invariant dilepton mass, somewhere in the
region $M\approx m_\rho $, we come to the double counting problem. This part
of the spectrum from the above reactions must be excluded 
as outlined below.

Experimental data on the exclusive cross sections $\sigma (s)_{P}^{VX}$ with 
$X=n\pi NN$ at $n\geq 1$ are not available. Here we assume that the
probabilities $w_{n}$ are described by a binomial distribution 
\begin{equation}
w_{n}=\frac{N_{\pi }!}{n!(N_{\pi }-n)!}p^{n}(1-p)^{N_{\pi }-n}.
\label{BINOM}
\end{equation}
To fix all probabilities it is sufficient to know the ratio between the
exclusive vector meson production cross section $\sigma (s)_{P}^{VNN}$ and
the inclusive cross section $\sigma (s)_{P}^{VX}$. These two cross sections
are experimentally known \cite{xsects}. The value $N_{\pi }$ is defined as
above by energy conservation while the value $p$ can be extracted from
the relation 
\begin{equation}
\frac{\sigma (s)_{P}^{VNN}}{\sigma (s)_{P}^{VX}}=(1-p)^{N_{\pi }}.  \label{p}
\end{equation}
The pion multiplicity equals 
\begin{equation}
n_{\pi }=\sum_{n=0}^{N_{\pi }}nw_{n}=pN_{\pi }.  \label{n_pi}
\end{equation}

In the case of the $\rho $-meson, the cross section $\sigma (s)_{P}^{VX}$
determines the pole behavior of the total cross section $d\sigma (s,M)^{\pi
^{+}\pi ^{-}X}$ in the vicinity of the $\rho $-meson peak. Like for vector
mesons, eq.(\ref{V2}), the total cross section $d\sigma (s,M)^{\pi ^{+}\pi
^{-}X}$ for the $2\pi $ production can be decomposed into $\rho $-pole and
background parts 
\begin{equation}
d\sigma (s,M)^{\pi ^{+}\pi ^{-}X}=d\sigma (s,M)_{P}^{\rho ^{0}X}+d\sigma
(s,M)_{B}^{\pi ^{+}\pi ^{-}X}.  \label{V5}
\end{equation}
 The background parts on the
right hand sides of eq.(\ref{V5}) and eq.(\ref{V2}) do not coincide
since the $\pi ^{+}\pi ^{-}$ quantum numbers are not necessarily equal to
the quantum numbers of the $\rho $-meson (this is the case at the $\rho $%
-meson peak only and this is why with $V=\rho ^{0}$ the first terms in eq.(%
\ref{V2}) and eq.(\ref{V5}) coincide). Thus the left hand sides
 of eq.(\ref{V5}) and eq.(\ref{V2}) do not coincide as well.
Since the $\rho $-meson is always
detected via $2\pi $ final states (or via $3\pi $ and $2K$ final states for $%
\omega $- and $\phi $-mesons, respectively), the inclusive cross section for
the production of vector mesons which enters into eq.(\ref{V1}) cannot be
uniquely determined from the experimental data on inclusive production
of $\pi ^{+}\pi ^{-}$ . Instead, one needs a model
for the calculation of the background part of the cross section in eq.(\ref
{V2}). A subtle problem of double counting in the total dilepton production
cross section appears in this way. We propose a quite natural phenomenological
solution for it.

The background term $d\sigma (s,M)_{B}^{\rho ^{0}X}$ at $X\neq NN\ $can be
saturated, at least partially, by considering the production of light
mesons: $pp\rightarrow \eta X\rightarrow $ $\rho ^{0}\gamma X\rightarrow \pi
^{+}\pi ^{-}\gamma X$, $pp\rightarrow \omega X\rightarrow \rho ^{0}\pi
^{0}X\rightarrow \pi ^{+}\pi ^{-}\pi ^{0}X$, etc., similarly for the $\omega 
$- and $\phi $-mesons. The $\pi ^{+}\pi ^{-}$ invariant masses are small
here, so these processes contribute to the $\pi ^{+}\pi ^{-}$ background.
Eq.(\ref{V1}) can now be rewritten as follows 
\begin{eqnarray}
d\sigma (s,M)^{e^{+}e^{-}X}&=&d\sigma (s,M)_{P}^{e^{+}e^{-}X}+d\sigma
(s,M)_{B}^{e^{+}e^{-}X}\left. {}\right| _{X=NN} \nonumber \\ &+&d\sigma
(s,M)_{B}^{e^{+}e^{-}X}\left. {}\right| _{X\neq NN}.  \label{XSEC}
\end{eqnarray}
The first term is the same as in Eq.(\ref{V1}), i.e. with the sum running
over all vector mesons, however, keeping only the pole part of the cross
section. In Eq.(\ref{XSEC}) the background is divided into contributions
from direct decays of intermediate vector mesons, which are off-shell and
typically below their physical thresholds ($X=NN$) and from decays of
intermediate mesons, ${\cal M}$, to multi-particle final states ($X\neq NN$%
). In the latter case, the cross section for the production of the
intermediate meson has to be folded over its branching ratio to the final
state under consideration:

\begin{eqnarray}
\frac{d\sigma (s,M)_{B}^{e^{+}e^{-}X}}{dM^{2}}\left. {}\right| _{X\neq
NN}&=&\sum_{{\cal M}}\int d\mu ^{2}(1+n_{{\cal M}}) \nonumber \\
&\times &\frac{d\sigma (s,\mu )^{%
{\cal M}X^{\prime }}}{d\mu ^{2}}\frac{dB(\mu ,M)^{{\cal M}\rightarrow e^{+}e%
{\normalsize X}^{\prime \prime }}}{dM^{2}}.  \label{o}
\end{eqnarray}
The sum runs over the mesons ${\cal M}=\pi ,\;\eta ,\;\rho ,\;\omega ,\;$and$%
\;\phi .$ Here, $n_{{\cal M}}$ is the average number of mesons ${\cal M}$ in
the state $X^{\prime }$. The value $\mu $ in the last equation describes the
distribution over the off-shell masses of the mesons. For pseudo-scalar
mesons, the cross sections due to their small widths are proportional to the
delta-function $\delta (\mu ^{2}-m_{{\cal M}}^{2})$ (cf. Eq.(\ref{zero-width}%
)) and the expression reduces to a sum over the on-shell mesons decaying to
the states $e^{+}eX^{\prime \prime }\ $with $X^{\prime \prime }\neq \oslash
, $ $X=X^{\prime }+X^{\prime \prime }$. For vector mesons entering the sum
of Eq.(\ref{o}) one should use the cross sections (\ref{V2}) whose pole
components are well defined.

The contribution to the background part of the cross section (\ref{V1}) with 
$X=NN\ $can be calculated assuming that it results from subthreshold decays
of baryon resonances $R=\Delta (1232),\;N(1520),\;...$ produced in $pp$
collisions, which decay into nucleons and vector mesons, $R\rightarrow NV$ 
\cite{BCM}. In terms of the branching ratios for the Dalitz decays of the
baryon resonances, the cross section can be written as follows

\begin{eqnarray}
\frac{d\sigma (s,M)_{B}^{e^{+}e^{-}X}}{dM^{2}}\left. {}\right|
_{X=NN}&=&\sum_{R}\int_{(m_{N}+M)^{2}}^{(\sqrt{s}-m_{N})^{2}}d\mu ^{2}\frac{%
d\sigma (s,\mu )^{pp\rightarrow pR}}{d\mu ^{2}} \nonumber \\
&\times &\sum_{V}\frac{dB(\mu
,M)^{R\rightarrow Vp\rightarrow e^{+}e^{-}p}}{dM^{2}}\cdot
\label{1}
\end{eqnarray}
Here, $\mu $ is the running mass of the baryon resonance $R$ with the cross
section $d\sigma (s,\mu )^{pp\rightarrow pR}$, $dB(\mu ,M)^{R\rightarrow
Vp\rightarrow e^{+}e^{-}p}$ is the differential branching ratio for the
Dalitz decay $R\rightarrow e^{+}e^{-}p$ through the vector meson $V$.

\begin{figure}
\begin{center}
\leavevmode
\epsfxsize = 13cm
\epsffile[40 40 530 450 ]{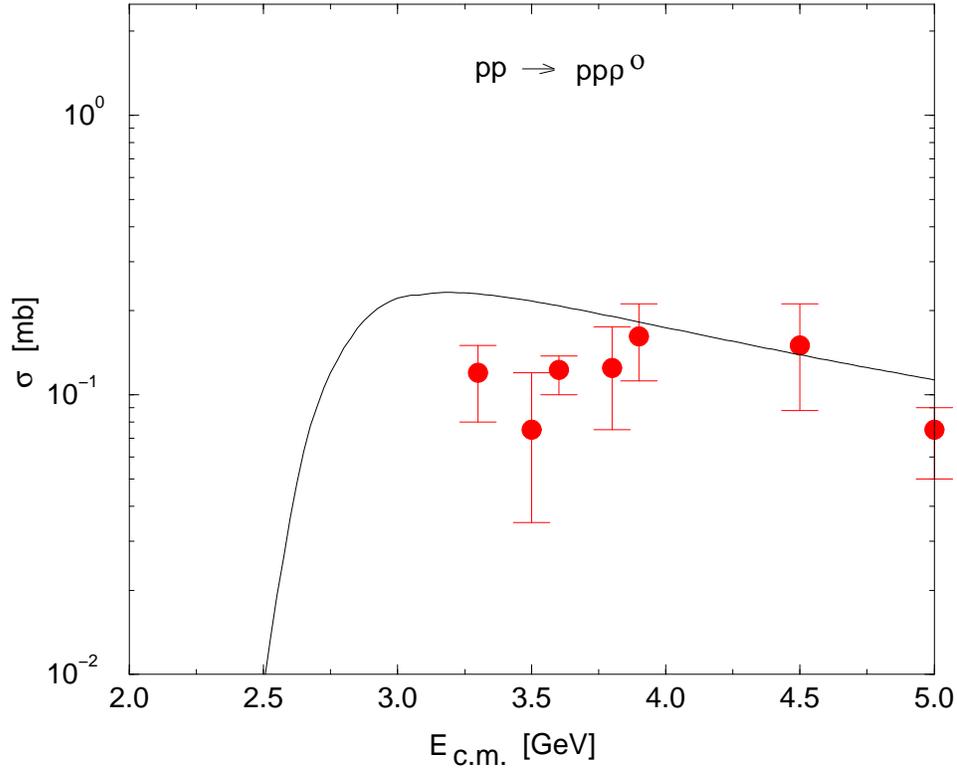}
\caption{\label{rho}Exclusive $\rho^0$ meson production in proton-proton
collision calculated over contribution of intermediate baryon
resonances (\ref{111}) compared to experimental data.}
\end{center}
\end{figure}

\begin{figure}
\begin{center}
\leavevmode
\epsfxsize = 13cm
\epsffile[40 40 530 450 ]{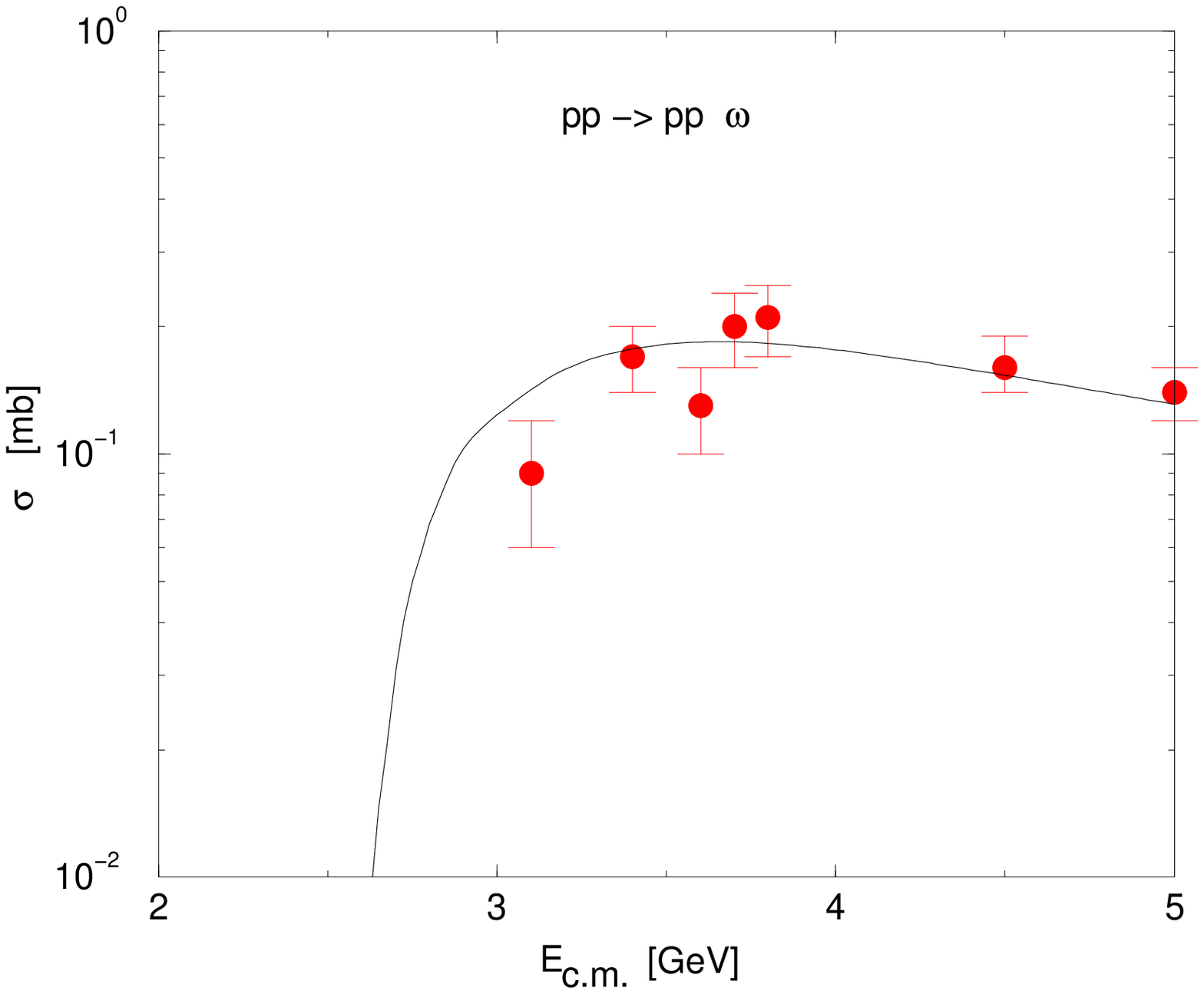}
\caption{Exclusive $\omega$ meson production in proton-proton
collision calculated over contribution of intermediate baryon
resonances (\ref{111}) compared to experimental data.
}
\end{center}
\label{omega}
\end{figure}

With increasing energy, the vector mesons in Eq.(\ref{1}) can be produced at 
their physical masses. In such a case, the  processes (\ref{1}) contribute
to the pole part of the cross section $d\sigma (s,M)^{e^{+}e^{-}X}$. The pole part of
inclusive cross sections by definition accounts for all possible sources for
the appearance of on-shell vector mesons, so that a naive extension of the
subthreshold cross section to higher energies would result in a double
counting. To avoid this double counting we should skip
the exclusive part of vector meson production in $pp$ collision that enters
to the total inclusive production of vector mesons as already taken into 
account in Eq.(\ref{1}):
\begin{equation}
d \sigma (s)=d \sigma (s)^{incl} (1-w_0) + d \sigma (s)^{subth}
\label{222}
\end{equation}
The factor $1-w_0$ excludes the $NNe^+e^-$ component of the inclusive cross 
section which is reproduced by the subthreshold mechanizm. 
This prescription lies on the
comparison of the experimental data to the exclusive vector meson production 
cross section, that goes through baryon resonances
\begin{eqnarray}
\sigma (s)^{VX}\left. {}\right|
_{X=NN}&=&\sum_{R}\int_{(m_{N}+M)^{2}}^{(\sqrt{s}-m_{N})^{2}}d\mu ^{2}\frac{%
d\sigma (s,\mu )^{pp\rightarrow pR}}{d\mu ^{2}} \nonumber \\
&\times &\int\frac{dB(\mu
,M)^{R\rightarrow Vp}}{dM^{2}}{dM^{2}}~.
\label{111}
\end{eqnarray}
The cross sections reasonably agree (see Figs. \ref{rho} and \ref{omega}).

The meson multiplicities $n_{{\cal M}}$ in Eq.(\ref{o}) are set equal to
zero except for the neutral pions. For pions, we assume $n_{\pi ^{0}}=\frac{1%
}{2}n_{\pi }$ where $n_{\pi }$ is given by Eq.(\ref{n_pi}). The factor $%
\frac{1}{2}$ appears statistically in the limit $n\rightarrow \infty $ from
charge conservation which implies that in every channel the total number of the
neutral-pion pairs $\pi ^{0}\pi ^{0}$ must be equal to the number of the
charged-pion $\pi ^{+}\pi ^{-}$ pairs.

There are no experimental data on the pion production inclusive cross
section $\sigma (s,\mu _{\pi })^{pp\rightarrow \pi ^{0}X}$ at $T=1\div 5$
GeV. There exist, however, experimental data on the two-pion production
cross sections $\sigma (s,\mu _{\pi })^{pp\rightarrow \pi \pi NN}$ \cite
{xsectspipi}. The parameter $p$ of the pion number distribution can
therefore be estimated from equation 
\begin{equation}
\frac{\sigma (s)^{\pi ^{0}\pi NN}}{\sigma (s)^{\pi ^{0}pp}}=\frac{pN_{\pi }}{%
1-p}.  \label{p_pion}
\end{equation}
It can further be used to find the total inclusive cross section $\sigma
(s,\mu _{\pi })^{pp\rightarrow \pi ^{0}X}$ using the same relation as 
in Eq.(\ref{p}). The results for the pion 
multiplicities produced in reactions in
accordance with different mesons ${\cal M}$ are summarized in Table 1 for
those values of kinetic energies, $T,$ at which the dilepton cross sections
are measured by the DLS\ collaboration. We give there also the cross
sections obtained from the interpolation and/or extrapolation between the
available experimental points. The accuracy of these estimates is rarely
better than $20\%$. For the pion, we give an estimate for the inclusive
cross section. Notice that it is in reasonable agreement with the
prediction from the UrQMD transport model \cite{ernst}.

For mesons ${\cal M} = \eta ,\;\rho
,\;\omega ,\ $and$\;\phi ,$ estimates for the cross sections $pp\rightarrow 
{\cal M}\pi NN$ are given as derived from the distribution (\ref{BINOM}).

\begin{table}
\caption{The maximum number of pions, $N_{\pi },$ the average pion multiplicities, $%
n_{\pi },$ and the cross sections $\sigma ^{{\cal M}\pi NN},$ $\sigma ^{%
{\cal M}NN},$ and $\sigma ^{{\cal M}X}$ for production of 
mesons ${\cal M} =\pi ^{0},\;\eta ,\;\rho ,\;\omega ,\ $and$\;\phi $ 
in the proton-proton
collisions for the set of energies $T=1.04$, $1.27$, $1.61$, $2.09$, and $%
4.88$ GeV, at which the dilepton production cross section has been 
measured by the DLS\ Collaboration. The last three lines with the cross sections contain 
$ 3 \times 12 = 36 $ numbers. The 24 numbers are 
compillation of the available experimental data, while the 12 numbers 
marked with 
the symbol $"^{\#}"$ are predictions from our binomial formula (\ref{BINOM}). 
The experimental cross sections are
measured at energies different from the BEVALAC energies, so we give 
interpolations from the available experimental
points.}
\lineup
\begin{tabular}{|c|c|c|c|c|c|c|}
\br
$\;\;T\;$[GeV]     & 1.04         & 1.27         & 1.61         & 1.85          & 1.85      & 2.09         \\ \mr
$\;\;\;\;{\cal M}$ & $\;\pi ^{0}$ & $\;\pi ^{0}$ & $\;\pi ^{0}$ & $\;\pi^{0} $  & $\;\eta $ & $\;\pi ^{0}$  \\ 
$\;\;\;\; N_{\pi }$& $\;2$        & $\;3$        & $\;3$        & $\;4$         & $\;1$     & $\;5$        \\ 
$\;\;\;\; n_{\pi }$& $0.09$       & $0.20$       & $0.56$       & $0.74$        & $0.26$    & $1.07$       \\ \mr
$\sigma ^{{\cal M}\pi NN}\;$[mb] 
                   & $0.4$        & $0.9$        & $2.7$        & $3.4$         & $4.9$     & $0.05^{\#)}$ \\ 
$\sigma ^{{\cal M}NN}\;\;$[mb] 
                   & $4.5$        & $4.2$        & $3.9$        & $3.8$         & $0.13$    & $3.6$       \\ 
$\sigma ^{{\cal M}X}\;\;\;\;$[mb] 
                   & $4.9^{\#)}$  & $5.2^{\#)}$  & $7.3^{\#)}$  & $8.5^{\#)}$   & $0.17$    & $12^{\#)}$   \\ \br
$\;\;T\;$[GeV]     & 2.09         & 4.88         & 4.88         & 4.88          & 4.88       & 4.88        \\ \mr
$\;\;\;\;{\cal M}$ & $\;\eta $    & $\;\pi ^{0}$ & $\;\eta $    & $\;\rho ^{0}$ & $\;\omega$ & $\;\phi $   \\ 
$\;\;\;\;N_{\pi }$ & $\;2$        & $\;11$       & $\;8$        & $\;6$         & $\;6$      & $\;4$       \\ 
$\;\;\;\;n_{\pi }$ & $0.55$       & $1.80$       & $1.87$       & $1.67$        & $1.36$     & $1.54$      \\ \mr
$\sigma ^{{\cal M}\pi NN}\;$[mb]  
                   & $0.11^{\#)}$ & $5.8$        & $0.34^{\#)}$ & $0.28^{\#)}$  &$0.35^{\#)}$& $0.003^{\#)}$\\ 
$\sigma ^{{\cal M}NN}\;\;$[mb] 
                   & $0.14$       & $2.7$        & $0.14$       & $0.12$        & $0.20$     & $0.001$      \\ 
$\sigma ^{{\cal M}X}\;\;\;\;$[mb] 
                   & $0.27$       & $19^{\#)}$   & $1.19$       & $0.94$        & $0.94$     & $0.007$       \\ \br

\end{tabular}
\end{table}


Also the estimates 
for the exclusive multi-pion production cross 
are in line with prediction from the FRITIOF string fragmentation 
model \cite{fritiof} and the calculation of ref.\cite{haglin93}. To
show this we make the following comparisons:

First,  we compare our simple 
formulas explicitely to FRITIOF. In refs.\cite {1,2} an attempt was made 
to extrapolate the FRITIOF model, originally proposed for high energies, to lower energies by
a modification of the two-body mechanism for inelastic hadron-hadron reactions. The
authors claimed that they got finally a reasonable description of the basic 
characteristics of the hadron-hadron and nucleus-nucleus collisions, 
within the modified FRITIOF model.

\begin{table}
\caption{The multipion production cross sections estimated in the FRITIOF model
(second line) and from the binomial distribution (third line) for the proton beam momentum 
of $P_{lab}=3.83$ and $5.1$ GeV. Given in the first line are the experimental data.}
\lineup
\begin{tabular}{|c|c|c|}
\br
$P_{lab}\;[GeV]$  & $3.83$ & $5.1$ \\ \mr
$np\rightarrow pp\pi^{-}$&$2.35 \pm 0.12$&$2.13 \pm 0.11$\\ 
$$&$(1.62)$&$(1.75)$\\ 
$$&$2.35$&$2.13$\\\hline
$np\rightarrow pp\pi^{-}\pi^{0}$&$1.83 \pm 0.13$ &$2.05 \pm 0.12$\\
$$&$(3.03)$&$(2.60)$\\
$$&$2.05$& $2.99$\\\hline
$np\rightarrow pp\pi^{-}\pi^{+}\pi^{-}$ &$0.31 \pm 0.04$&$0.56 \pm 0.04$\\
$$&$(0.41)$&$(0.90)$\\
$$&$0.38$&$0.93$\\\hline
$np\rightarrow pp\pi^{-}\pi^{0}\pi^{+}\pi^{-}$&$0.08 \pm 0.01$&$0.34 \pm 0.03$\\ 
$$&$(0.06)$&$(0.28)$\\ 
$$& $0.08 $ & $0.34$\\  \br
\end{tabular}
\end{table}

We compare our model with the modified FRITIOF model of refs.\cite{1,2}
for the reaction $np\rightarrow pp\pi^{-}X$ at $P_{lab}$ = 3.83 GeV and 5.1 GeV. 
In Table 2, we show
4 channels with $X = \oslash, \pi^{0}, \pi^{+}\pi^{-}, \pi^{+}\pi^{-}\pi^{0}$. 
The cross sections for the channels $X = \pi^{0}\pi^{0}, \pi^{0}\pi^{0}\pi^{0}$ 
are not available. We assume
\begin{equation}
\sigma(np \rightarrow pp\pi^{-}\pi^{0}\pi^{0}) 
= \sigma(np\rightarrow pp\pi^{-}\pi^{+}\pi^{-}), 
\label{XX}
\end{equation}
\begin{equation}
\sigma(np\rightarrow pp\pi^{-}\pi^{0}\pi^{0}\pi^{0}) 
= \sigma(np\rightarrow pp\pi^{-}\pi^{+}\pi^{-}\pi^{0}). 
\label{YY}
\end{equation}
We use the 
experimental data from the first and the last lines of the Table 2 to fit 
the parameter $p$ of the binomial distribution. 
The results are shown in the Table 2 (third lines in the boxes), together 
with the experimental data (first lines of the boxes, numbers with errors) 
and predictions of the modified FRITIOF model (second lines of the 
boxes, numbers in the brackets). The experimental data are from ref. \cite{3}.
Is is seen that the description is reasonable. Notice that a $30\%$ accuracy 
occur for the modified FRITIOF predictions and similarly $30\%$ is also the accuracy
of the binomial formula predictions.

Second, the agreement with the work \cite{haglin93} is reasonable. 
For example as it can be seen from Table 2 that our value for 
$\sigma(np\rightarrow pp\pi^{-}\pi^{0})$ at 5.1 GeV is close to 
their result of 2.85 mb at 5.5 GeV.

In Table 3, we compare predictions from the binomial distribution for
the reaction $pp \rightarrow pp\pi^{0}X$ with
the experimental data from ref.\cite{Alex}.
The uncertainty exists in these data also: the cross sections for $X = \pi^{0}$, 
$\pi^{0}\pi^{0}$, $\pi^{0}\pi^{0}\pi^{0}$,
$\pi^{0}\pi^{+}\pi^{-}$, $\pi^{0}\pi^{0}\pi^{0}\pi^{0}$, 
and $\pi^{0}\pi^{0}\pi^{-}\pi^{+}$ are unknown. For a qualitative
estimate, we assume like before that the cross sections for the unknown
channels are equal to the cross sections for the known channels, with the same number
of the pions. The parameter $p$ of the binomial formula is determined at each energy
from the channels with one pion and with the maximum number of the pions in the final states. 

The agreement is not unreasonable, especially in the view of the naive approximations 
like (\ref{XX}) and (\ref{YY}). The binomial formula does not contradict to the available 
data and we suppose that it can be used for estimates of the multiple pion production 
effects. Notice that the pion contribution to the dilepton spectrum is quite far from the 
really interesting region about and just below the $\rho$-meson peak.

For the mesonic dilepton decays$\;$of mesons ${\cal M}\rightarrow
e^{+}e^{-}X,$ experimental data exist in most cases. The radiative
decays ${\cal M}\rightarrow \gamma X$ can further be used to calculate the
dilepton decays when the experimental data are not available. For details of
the determination of the various branching ratios see ref.\cite{kriv}. The
inclusive cross sections for the meson production and the branching ratios
for the mesonic decays to dileptons can be combined to estimate the dilepton
yield from Eq.(\ref{XSEC}).

\begin{table}
\caption{The multipion production cross sections in proton-proton collisions
estimated form the binomial distribution for different momenta of the beam proton (first 
coulomn). The second coulomn shows the final states, the third one shows the parameter 
$p$ of the binomial distribution, the next coulomn gives the estimated cross sections 
in $mb$, and the last one gives the experimental numbers in $mb$.}

\lineup
\begin{tabular}{|c|c|c|c|c|}
\br
$P_{lab}\;[GeV]$  & $final\;state$   & $p$ & $\sigma^{binom}\;[mb]$ & $\sigma^{expt}\;[mb]$ \\ \hline
$5.5  $ & $pp\pi^{0}$                   & $0.127$   &  $2.77   $  &  $  2.77   \pm   0.11$     \\ 
$5.5  $ & $pp\pi^{+}\pi^{-}$            & $0.127$   &  $4.04   $  &  $  2.84   \pm   0.08$     \\ 
$5.5  $ & $pp\pi^{+}\pi^{-}\pi^{0}$     & $0.127$   &  $1.327 $  &  $  1.81   \pm   0.07$     \\ 
$5.5  $ & $pp\pi^{+}\pi^{-}\pi^{+}\pi^{-}$      & $0.127$   &  $ 0.516$  &  $  0.227 \pm   0.023$ \\ 
$5.5  $ & $pp\pi^{+}\pi^{-}\pi^{+}\pi^{+}\pi^{0}$       & $0.127$   &  $ 0.088$  &  $  0.088 \pm   0.014$   \\ 
$3.68$ & $pp\pi^{0}$            & $0.134$   &  $ 2.950$  &  $  2.95   \pm   0.31$     \\
$3.68$ & $pp\pi^{+}\pi^{-}$             & $0.134$   &  $ 3.213$  &  $  2.72   \pm   0.13$     \\ 
$3.68$ & $pp\pi^{+}\pi^{-}\pi^{0}$      & $0.134$   &  $ 0.750$  &  $  0.75   \pm   0.07$     \\ 
$2.81$ & $pp\pi^{0}$            & $0.095$   &  $ 3.600$  &  $  3.60   \pm   0.21$     \\ 
$2.81$ & $pp\pi^{+}\pi^{-}$             & $0.095$   &  $ 1.897$  &  $  2.35   \pm   0.14$     \\ 
$2.81$ & $pp\pi^{+}\pi^{-}\pi^{0}$      & $0.095$   &  $ 0.200$  &  $  0.20   \pm   0.03$     \\ 
$2.23$ & $pp\pi^{0}$            & $0.054$   &  $ 4.060$  &  $  4.06   \pm   0.27$     \\ 
$2.23$ & $pp\pi^{+}\pi^{-}$             & $0.054$   &  $ 0.697$  &  $  1.24   \pm   0.14$     \\ 
$2.23$ & $pp\pi^{+}\pi^{-}\pi^{0}$      & $0.054$   &  $ 0.020$  &  $  0.02   \pm   0.02$     \\ \br
\end{tabular}
\end{table}

The differential decay branchings $dB^{R\rightarrow Ne^{+}e^{-}}(\mu
,M)/dM^{2}\;$are calculated in ref.\cite{BCM} in a non-relativistic
approximation for the multipole decays with the emission of a massive vector
particle. We follow a similar approach here, but
consider the relativistic case and modify the
transition form factors for the nucleon resonances, $R$, which is needed to bring
their asymptotic behavior in the correspondence with the quark counting rules
and to provide an unified description of the photo- and electroproduction data 
and vector meson decays $R\rightarrow N\rho (\omega )$ \cite{FFKM}.

\section{Transition form factors, quark counting rules, and radiative and
vector meson decays of nucleon resonances}

The problem is adequately formulated in the non-relativistic
approximation for radiative and vector meson decays of nucleon resonances. 
We start with the discussion of this case, since it is much simpler.
The relativistic treatment, which will finally be
implemented into the consideration, although is more complicated,
is motivated by the same physical ideas.

\subsection{Vector meson decays of nucleon resonances in the nonrelativistic 
approximation}

The description of the resonance decays $R\rightarrow N\gamma ^{*},$ $\gamma
^{*}\rightarrow e^{+}e^{-}$ is usually based on the VMD model which provides
transition form factors $RN\gamma $ of a monopole form. The pole corresponds
to the masses of the $\rho $- and $\omega $-mesons. This model should give, in
principle, an unified description of the radiative $RN\gamma $ and the mesonic 
$RNV$ decays. However, a normalization to the radiative branchings ($%
RN\gamma $) strongly underestimates the mesonic branchings ($RNV$) as we
discuss below.

The resonance $N(1520)$ is a case for which both, the $N(1520)\rightarrow
N\rho $ and $N(1520)\rightarrow N\gamma $ widths are known with a relatively
high precision: $B(N(1520)\rightarrow N\rho )=15\div 25\%$, $%
B(N(1520)\rightarrow N\gamma )=0.46\div 0.56$ $\%$ ($p\gamma $ mode)$,$ $%
0.30\div 0.53$ $\%$ ($n\gamma $ mode). The branching ratios of the proton
and neutron modes are equal within the experimental errors. Let us, 
for the moment, interpreete this by assumption that the radiative mode is dominated by the $%
\rho $-meson. The same conclusion is reasonable for
other $N^*$ resonances: $B(N(1440)\rightarrow N\gamma )=0.035\div 0.048$ $\%$
($p\gamma $ mode)$,$ $0.009\div 0.032$ $\%$ ($n\gamma $ mode); $%
B(N(1535)\rightarrow N\gamma )=0.15\div 0.35$ $\%$ ($p\gamma $ mode)$,$ $%
0.004\div 0.29$ $\%$ ($n\gamma $ mode), etc. The $\Delta $ decays
 proceed exclusively through the $\rho $-meson.

However, now the standard VMD model as it has been used in \cite{BCM} leads
to a severe inconsistency: Using the coupling constant $f_{N(1520)N\rho
}=7.0 $ extracted from the mesonic $N(1520)\rightarrow N\rho $ decay, the
branching ratio for the radiative decay is found to be two to three
times greater than the experimental value. Analogous overestimations are
observed almost for all other $N$ and $\Delta $ resonances for which the
experimental $N\rho $ and $N\gamma $ data are available. Table 4 summarizes
the results.

\begin{table}
\caption{The coupling constants $f_{RN\rho }$ derived from the $R\rightarrow
N\rho $ mesonic decays are compared to the coupling constants $f_{RN\rho
}^{\gamma }$ fixed from the radiative $R\rightarrow N\gamma $ decays. The
numerical values $f_{RN\rho }$ are taken from ref. \cite{PPLLM}, with
exception of the $\Delta (1232)$ resonance for which the theoretical value
from \cite{RAPP} is given and of the $N(1440)$ and $N(1535)$ resonances
where the results of our calculations are given.}

\begin{tabular}{|c|c|c|c|c|c|c|c|c|c|c|}
\hline\hline
$R$ & $N_{1440}$ & $N_{1520}$ & $N_{1535}$ & $N_{1650}$ & $N_{1680}$ & $N_{1720}$
& $\Delta _{1232}$ & $\Delta _{1620}$ & $\Delta _{1700}$ & $\Delta _{1905}$ \\ 
\hline\hline
$J^{P}$ & $\frac{1}{2}^{+}$ & $\frac{3}{2}^{-}$ & $\frac{1}{2}^{-}$ & $\frac{%
1}{2}^{-}$ & $\frac{5}{2}^{+}$ & $\frac{3}{2}^{+}$ & $\frac{3}{2}^{+}$ & $%
\frac{1}{2}^{-}$ & $\frac{3}{2}^{-}$ & $\frac{5}{2}^{+}$ \\ \hline\hline
$f_{RN\rho }$ & $<$ 26 & 7.0 & $<$ 2.0 & 0.9 & 6.3 & 7.8 & 15.3 & 2.5 & 5.0
& 12.2 \\ \hline
$f_{RN\rho }^{\gamma }$ & 1.3 & 3.8 & 1.8 & $<$ 0.8 & 3.9 & 2.2 & 10.8 & 0.7
& 2.7 & 2.1 \\ \hline\hline
\end{tabular}
\end{table}

The standard VMD predicts a $1/t$ asymptotic behavior for the transition
form factors. However, quark counting rules require a stronger suppression
at high $t$. It is known from the nucleon form factors, the pion form
factor, and the $\omega \pi \gamma $ and $\rho \pi \gamma $ transition form
factors that the quark counting rules start to work experimentally at
moderate $t\sim 1$ GeV$^{2}.$ One can assume that an appropriate
modification of the standard VMD which takes the correct asymptotics of the $%
RN\gamma $ transition form factors into account can provide a more accurate
description of the radiative decays of the nucleon resonances.

We propose the following solution of the inconsistency between the $RNV$ and 
$RN\gamma $ decay rates: Let radial excitations of the $\rho $-meson, the $%
\rho (1450)$-meson and $\rho (1700)$-meson, for example, interfere with the $\rho $-meson
in radiative processes. However, we know neither the couplings of the $\rho
(1450)$ and $\rho (1700)$ to the resonances $(f_{RN\rho ^{\prime }},$ $%
f_{RN\rho ^{\prime \prime }}$) nor the couplings of the $\rho (1450)$ and $%
\rho (1700)$ to a photon ($g_{\rho ^{\prime }},$ $g_{\rho ^{\prime \prime }}$%
). Thus in the sum 
\begin{eqnarray}
{\cal M}(M^{2})=\sum_{i=1}^{3}{\cal M}_{i}&=&\frac{f_{RN\rho }}{m_{\rho }}\frac{m_{\rho }^{2}%
}{g_{\rho }}\frac{1}{\tilde{m}_{\rho }^{2}-M^{2}}+\frac{f_{RN\rho ^{\prime }}%
}{m_{\rho ^{\prime }}}\frac{m_{\rho ^{\prime }}^{2}}{g_{\rho ^{\prime }}}%
\frac{1}{\tilde{m}_{\rho ^{\prime }}^{2}-M^{2}} \nonumber \\
&+&\frac{f_{RN\rho ^{\prime
\prime }}}{m_{\rho ^{\prime \prime }}}\frac{m_{\rho ^{\prime \prime }}^{2}}{%
g_{\rho ^{\prime \prime }}}\frac{1}{\tilde{m}_{\rho ^{\prime \prime
}}^{2}-M^{2}}~,  \nonumber
\end{eqnarray}
where 
$\tilde{m}_{k}^{2}=m_{k}^{2}-iM{\Gamma }_{k}\ $with$~k=\rho ,$ $%
\rho ^{\prime },$ and $\rho ^{\prime \prime }$,
$\rho ^{\prime }$ and $\rho ^{\prime \prime }$ refer to $\rho (1450)$-
and $\rho (1700)$-mesons, respectively, the coefficients $\frac{f_{RN\rho
^{\prime }}}{m_{\rho ^{\prime }}}\frac{m_{\rho ^{\prime }}^{2}}{g_{\rho
^{\prime }}}$ and $\frac{f_{RN\rho ^{\prime \prime }}}{m_{\rho ^{\prime
\prime }}}\frac{m_{\rho ^{\prime \prime }}^{2}}{g_{\rho ^{\prime \prime }}}$
are unknown. According to the quark
counting rules \cite{QCR,VZ}, for large and negative $M^{2}$ the form
factors of the $RN\gamma ^{*}$ amplitudes decrease like $1/M^{6}$. On the
phenomenological level we can attribute such a behavior to a cancellation
between the $\rho $-$,$ $\rho ^{\prime }$-$,$ and $\rho ^{\prime \prime }$%
-mesons. The constants $\frac{f_{RN\rho ^{\prime }}}{m_{\rho ^{\prime }}}%
\frac{m_{\rho ^{\prime }}^{2}}{g_{\rho ^{\prime }}}$ and $\frac{f_{RN\rho
^{\prime \prime }}}{m_{\rho ^{\prime \prime }}}\frac{m_{\rho ^{\prime \prime
}}^{2}}{g_{\rho ^{\prime \prime }}}$ are then fixed and we obtain 
\begin{equation}
{\cal M}(M^{2})=\frac{f_{RN\rho }}{m_{\rho }}\frac{m_{\rho }^{2}}{g_{\rho }}\frac{1%
}{\tilde{m}_{\rho }^{2}-M^{2}}\left( \frac{\tilde{m}_{\rho ^{\prime }}^{2}-%
\tilde{m}_{\rho }^{2}}{\tilde{m}_{\rho ^{\prime }}^{2}-M^{2}}\right) \left( 
\frac{\tilde{m}_{\rho ^{\prime \prime }}^{2}-\tilde{m}_{\rho }^{2}}{\tilde{m}%
_{\rho ^{\prime \prime }}^{2}-M^{2}}\right) .  \label{8}
\end{equation}
The last two factors in Eq.(\ref{8}) give the desired modification of the $%
\rho $-meson contribution to the radiative decays of the baryon resonances,
as compared to the naive VMD\ model: 
\begin{equation}
d\Gamma ^{(R\rightarrow Ne^{+}e^{-})}(\mu ,M)=d\Gamma ^{(R\rightarrow
Ne^{+}e^{-})}(\mu ,M)^{(naive\;VMD)}F_{\rho }(M^{2}).  \label{9}
\end{equation}
The mass-dependent correction factor is given by 
\begin{equation}
F_{\rho }(M^{2})=\left| \left( \frac{\tilde{m}_{\rho ^{\prime }}^{2}-\tilde{m%
}_{\rho }^{2}}{\tilde{m}_{\rho ^{\prime }}^{2}-M^{2}}\right) \left( \frac{%
\tilde{m}_{\rho ^{\prime \prime }}^{2}-\tilde{m}_{\rho }^{2}}{\tilde{m}%
_{\rho ^{\prime \prime }}^{2}-M^{2}}\right) \right| ^{2}.
\end{equation}
The same modification applies to the $R\rightarrow N\gamma $ decays. The
reduction factor in the amplitude $R\rightarrow N\gamma $ equals $\sqrt{%
F_{\rho }(M^{2}=0)}=0.56$. It is seen from Table 4 that a reduction of about 
$\frac{1}{2}$ is just what one needs for a consistent description of both,
the $\rho $-meson and the radiative decay of the $N(1520)$. In all other
cases the reduction factor also improves the agreement. In the case of the $\Delta
(1905)$ resonance, the large difference between $f_{RN\rho }$ and $f_{RN\rho
}^{\gamma }$ can be attributed to a further suppression of the amplitude $%
A(M^{2})$ due to the quark counting rules which require a $1/M^{8}$ behavior
of the $RN\gamma ^{*}$ vertex for the $\frac{5}{2}^{+}\rightarrow \frac{1}{2}%
^{+}$ transition. So, we have formulated the problem and outlined its possible solution.

Note that interference between the different $\rho$-meson states does not change
essentialy the dilepton contribution from the pion annihilation channel in heavy-ion 
collisions, since the $\rho$-meson form factor involved into that process falls off 
asymptotically like $1/t$ due to the quark counting rules, and so there should be no 
destructive interference between members of the $\rho$-meson family. The VMD model 
by Kroll, Lee and Zumino \cite{KLZ} allows two independent couplings for the photons 
and vector mesons and can be used to resolve the discrepancy between the photon and
$\rho$-meson branchings of the nucleon resonances \cite{FP}. It provides, however, 
the form factors asymptotics $F(t) = O(1)$ in disagreement with the quark counting rules.

\subsection{Relativistic treatment of the vector meson decays of nucleon resonances}

The relativistic treatment of the $R \rightarrow NV$ decays is in details discussed in [37].
Here, we sketch out the basic concept.
For nucleon resonances with spin $J>1/2$ and arbitrary parity, 
there exist three independent transition form factors, while 
for spin-1/2 resonances, two
independent form factors should be considered \cite{JS,DEK,Trueman:1969wn}. 

In terms of the electric (E), magnetic (M), and Coulomb (C) form factors, the decay widths
of nucleon resonances with spin $J = l + 1/2$ into a virtual photon with mass $M$ 
has the form \cite{Krivoruchenko:2001hs,FFKM}:

\begin{eqnarray}
\Gamma (N_{(\pm )}^{*} &\rightarrow &N\gamma ^{*})=\frac{9\alpha }{16}\frac{%
(l!)^{2}}{2^{l}(2l+1)!}\frac{m_{\pm }^{2}(m_{\mp }^{2}-M^{2})^{l+1/2}(m_{\pm
}^{2}-M^{2})^{l-1/2}}{m_{*}^{2l+1}m^{2}}  \nonumber \\
&&\left( \frac{l+1}{l}\left| G_{M/E}^{(\pm )}\right| ^{2}+(l+1)(l+2)\left|
G_{E/M}^{(\pm )}\right| ^{2}+\frac{M^{2}}{m_{*}^{2}}\left| G_{C}^{(\pm
)}\right| ^{2}\right),  \label{GAMMA_l}
\end{eqnarray}
where $m_*$ refers to the nucleon resonance mass, $m$ is the nucleon mass,
$m_{\pm} = m_* \pm m$. The signs $\pm$ refer to the natural parity ($1/2^-,
3/2^+, 5/2^-,$ ...) and abnormal parity ($1/2^+,
3/2^-, 5/2^+,$ ...) resonances. $G_{M/E}^{\pm}$ means $G_{M}^{+}$ or $G_{E}^{-}$. 
The above equation is valid
for $l>0$. For $l=0$ ($J=1/2$), one gets 

\begin{eqnarray}
\Gamma (N_{(\pm )}^{*} &\rightarrow &N\gamma ^{*})=\frac{\alpha }{8m_{*}}%
(m_{\pm }^{2}-M^{2})^{3/2}(m_{\mp }^{2}-M^{2})^{1/2}  \nonumber \\
&&\left( 2\left| G_{E/M}^{(\pm )}\right| ^{2}+\frac{M^{2}}{m_{*}^{2}}\left|
G_{C}^{(\pm )}\right| ^{2}\right).  \label{GAMMA_0}
\end{eqnarray}
\begin{figure}
\begin{center}
\leavevmode
\epsfxsize = 13cm
\epsffile[40 40 530 450 ]{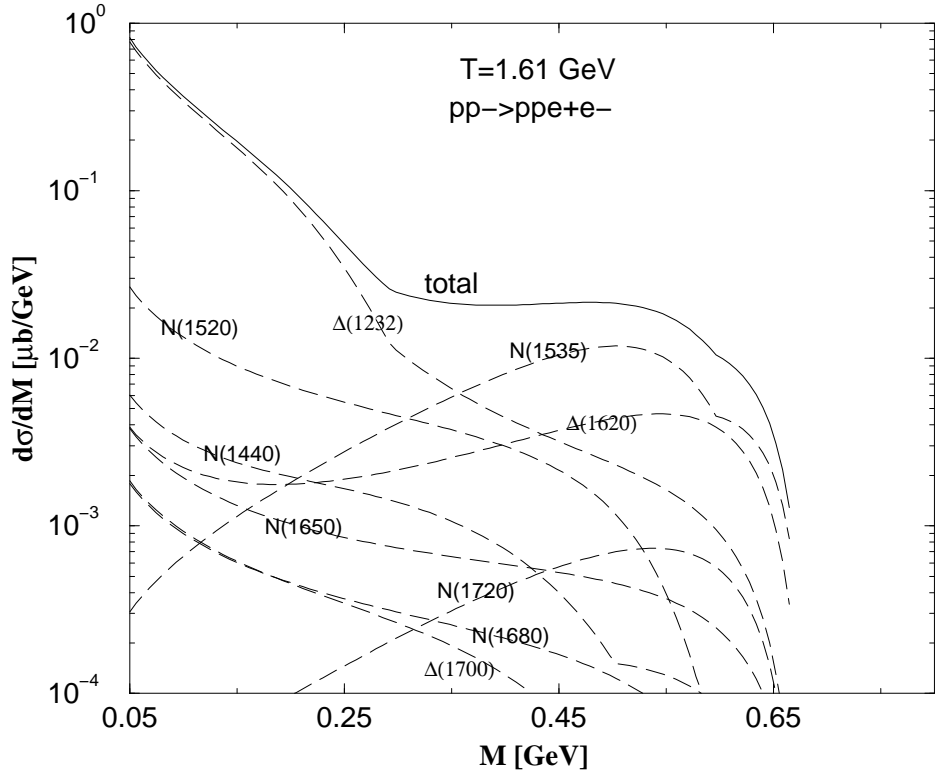}
\caption{The dilepton production cross sections 
$pp\rightarrow e^{+}e^{-}pp$
through the nucleon resonances $R=\Delta ,$ $N^{*},$ and $\Delta ^{*}$ at
an kinetic proton energy of $T=1.61$ GeV.}
\end{center}
\label{figx}
\end{figure}

If the width $\Gamma (N^{*}\rightarrow N\gamma ^{*})$ is known, the
factorization prescription can be
used to find the dilepton decay rate:

\begin{equation}
d\Gamma (N^{*}\rightarrow Ne^{+}e^{-})=\Gamma (N^{*}\rightarrow N\gamma
^{*})M\Gamma (\gamma ^{*}\rightarrow e^{+}e^{-})\frac{dM^{2}}{\pi M^{4}},
\label{OK!}
\end{equation}
where 
\begin{equation}
M\Gamma (\gamma ^{*}\rightarrow e^{+}e^{-})=\frac{\alpha }{3}%
(M^{2}+2m_{e}^{2})\sqrt{1-\frac{4m_{e}^{2}}{M^{2}}}  \label{OK!!}
\end{equation}
is the decay width of a virtual photon $\gamma ^{*}$ into the dilepton
pair with invariant mass $M$.

The couplings of the $\rho$- and $\omega$- mesons and of their radial exitations 
can be taken into account to describe radiative and electroproduction helicity 
amplitudes satisfying the quark counting rules. The available data on the
partial-wave analysis of the multichannel $\pi N$-scattering can also be included
into the framework of the extended VMD model. In ref.\cite{FFKM}, the 
parameters of the model are fixed by fitting those data and by taking into account
quark model predictions when the experimental data are not available. 
In this work, we use the model \cite{FFKM} to calculate the dilepton production from the nucleon 
resonance decays. 

In the relativistic case, we cannot write a compact expression for the 
suppression factor (23). However, it is clear that the effect does exist.
The destructive interference is prescribed by the quark counting 
rules which are implemented into the relativistic 
model. It means that the radiative decay should be less probable as compared
to the naive VMD estimate from the ground-state $\rho$-meson. 

The medium can destroy the destructive interference, in which case one can 
expect an enhancement of the dilepton production below the $\rho$-meson peak 
\cite{FFKM}. A detailed treatment of this effect will be given elsewhere \cite{SFFKM}. 

The $\Delta (1232)$ resonance is treated in the same way as the other
resonances. We take into acocunt $10$ resonances listed in Table 4. 
The $\rho$- and $\omega$-meson channels are treated on the same footing. The
numerical results demonstrate that besides the $N(1520)$ and $\Delta (1232)$
resonances, the $N(1535)$ and $\Delta (1620)$ have considerable
contributions. It can be seen from Fig. 3 where the resonance
contributions are shown for proton kinetic energy $T=1.61~$GeV.
At moderate invariant masses $M\leq 0.35~$GeV of the
dilepton pair, the resonance contributions are dominated by the $\Delta
(1232)$. At larger masses $M\geq 0.35~$GeV, contributions from the heavier
resonances become dominant. 


\section{ Numerical results}

\begin{figure}
\begin{center}
\leavevmode
\epsfxsize = 12cm
\epsffile[10 40 560 640]{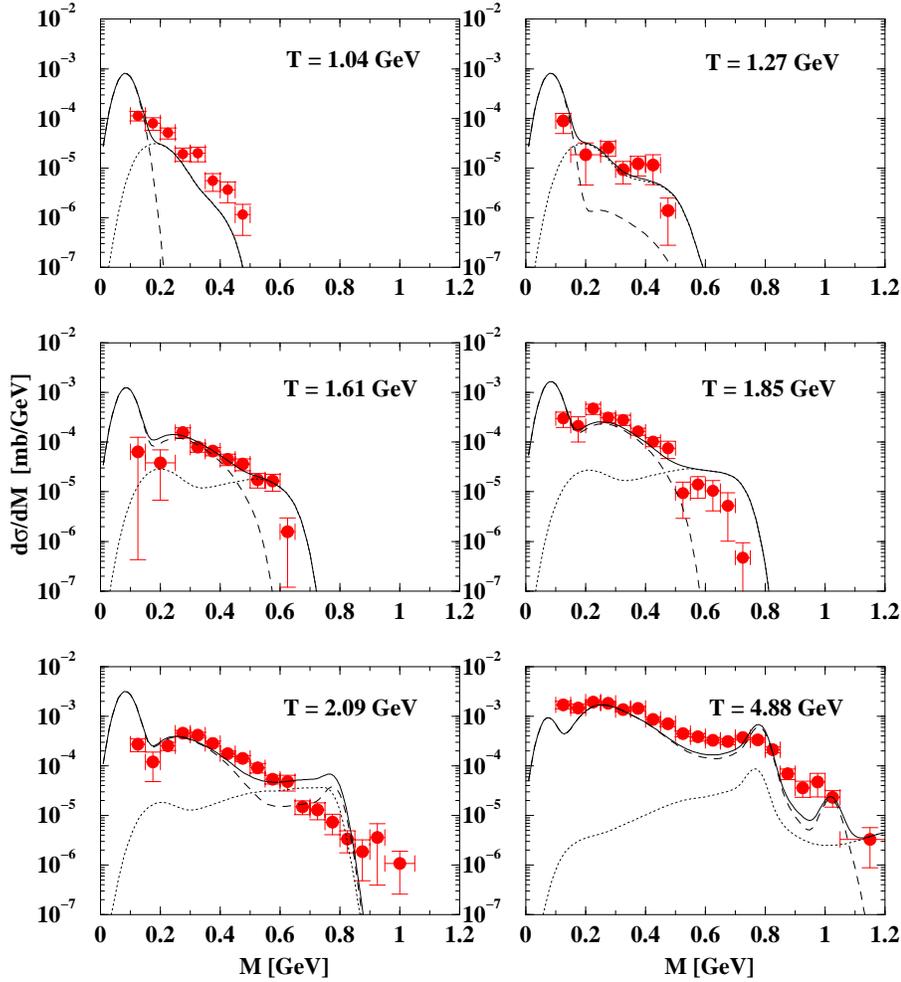}
\end{center}
\caption{The differential dilepton production cross sections as a
function of the dilepton invariant mass, $M$, after applying the
experimental filter and the smearing procedure (see text). The solid
curves are the total cross sections, the dashed curves correspond to the
inclusive production, and the dotted curves correspond to the subthreshold
production. The experimental data are from ref.\cite{BEVALAC}.
}
\label{fig3}
\end{figure}

The results for the dilepton spectra are shown in Fig. \ref{fig3}. We show 
also inclusive and subthreshold cross sections separately. To compare with
the experimental data, the acceptance of the DLS\ detector with respect to
the $e^{+}e^{-}$ pairs that have invariant mass $M$, transverse momentum $%
p_{T}$, and rapidity $y$ is taken into account. For each process, the
distribution over the $p_{T}$ and $y$ is determined by the available phase
space of the process and then weighted with the filter function $%
f(M,p_{T},y) $ provided by the DLS collaboration. 
The details of this procedure
are described in Appendix B. Finally, the finite mass resolution of the
detector, $\Delta M^{expt}=\pm \;25$~MeV, is taken into account by smearing
the spectra with a Gaussian distribution which corresponds to a standard
error of $\sigma =25~$MeV.

At the lowest initial kinetic energy of the proton, i.e. $T=1.04~$GeV, the
cross section is dominated by the $\pi ^{0}$-Dalitz decay below $M\leq
100~$MeV and by the Dalitz decays of the nucleon resonances, mainly the $%
\Delta (1232)$-resonance, at $M\approx 200\div 500~$MeV. Compared to our
calculation there is an excess of detected $e^{+}e^{-}$ pairs at $M\geq
300$ MeV. Earlier calculations \cite{BCM} obtained higher cross sections in
this mass range. This is due to the normalization to the $R\rightarrow N\rho 
$ branching ratios within the framework of the naive VMD which
overestimates the radiative decay rates $R\rightarrow N\gamma $, as
discussed in Sect.3. 

At higher initial proton energies, the agreement with the DLS\ data is
generally very reasonable. The contribution from the $\eta $-meson Dalitz
decay is dominant at $M\approx 0.2-0.4~$GeV, while the Dalitz decays of the
baryon resonances dominate at $M\geq 400~$MeV.

For the proton kinetic energy $T=2.09~$GeV, the inclusive and subthreshold
productions of the $\rho $- and $\omega $-mesons become important at 
$M\approx 800~$MeV. The $\omega $-meson peak in the inclusive cross section
is rather pronounced, whereas the subthreshold $\omega $-meson production 
cross section is large and smooth. The total cross section, as a result,
exceeds significantly the experimental data at $M\approx 0.7-0.8~$GeV.

Refs.\cite{ernst,BCM} agree quite well with the experimental data at the 
$\omega$-meson region. In ref.\cite{ernst}, the inclusive $\omega$-peak 
is not reproduced, apparently, due to a stronger {\it ad hoc} smearing 
with a mass-dependent parameter $\sigma =0.1M$. The smearing at the 
$\omega$-meson peak turns out to be factor of 3 greater than it should 
($\sigma$ = 80 MeV instead of 25 MeV). In refs.\cite{ernst,BCM}, the 
$\omega$-chanels from the nucleon resonance decays are neglected. This is 
the reason for the difference between our results and relults of refs.\cite{ernst,BCM}. 
In our calculations, the $\omega$-chanel contribution is large. It can only be
lowered by price of an {\it ad hoc} reduction of the decay probabilities of the 
nucleon resonances into the vector mesons, which has been calculated using the quark models \cite{QM}.
The quark models 
reproduce, however, the $\rho$-meson decays of the nucleon resonances quite well, 
i.e. they are in good agreement with the partial-wave analysis of the $\pi N$ inelastic scattering 
\cite{Manley:1992yb}. In this way the experimental underestimation of the dilepton
yield at the $\omega$-peak at $T=2.09~$GeV seems to contradict the 
$\pi N$ inelastic scattering data.

Remarkably, the three
highest experimental points at $T=2.09~$GeV lie above the kinematical limit $%
M_{max}=\sqrt{s}-2m_{N}\approx 850~$MeV and the indicated experimental
error $\Delta M^{expt}=\pm \;25$~MeV is not sufficient to explain their
occurrence. We suppose that the experimental resolution is not good enough 
to resolve the kinematic threshold. 

Finally, at $T=4.88~$GeV, the contribution from the inclusive production of
the $\eta $-, $\rho $-, and $\omega $-mesons becomes dominant at $M\approx
300\div 800~$MeV. There is an underestimation of the dilepton yield in the
region $M\approx 400\div 700~$MeV. A similar underestimation was found in 
\cite{ernst} both, at $T=2.09~$GeV and $T=4.88~$GeV. As proposed in ref.\cite
{BCM}, the existing gap might be filled by the subthreshold dilepton
production via the baryon resonances. However, we were not able to match the
data using a consistent description of the photoproduction data and the $%
R\rightarrow N\rho $ meson decay branchings, with the proper application of
the Breit-Wigner formula (see details in Appendix 1), and removing possible
sources for the double counting. Each of this three aspects leads to a
reduction of the dilepton yield. Therefore, one cannot exclude that the
origin of the so called ''DLS puzzle'' can be traced back to the elementary $%
pp$ level and is not a specific feature of heavy-ion collisions. New
experimental measurements of the dilepton cross section, especially at 
$T$ = 1.04, 2.09, and 4.88 GeV, would certainly help to clarify this point.

\begin{figure}
\begin{center}
\leavevmode
\epsfxsize = 12cm
\epsffile[70 30 430 600]{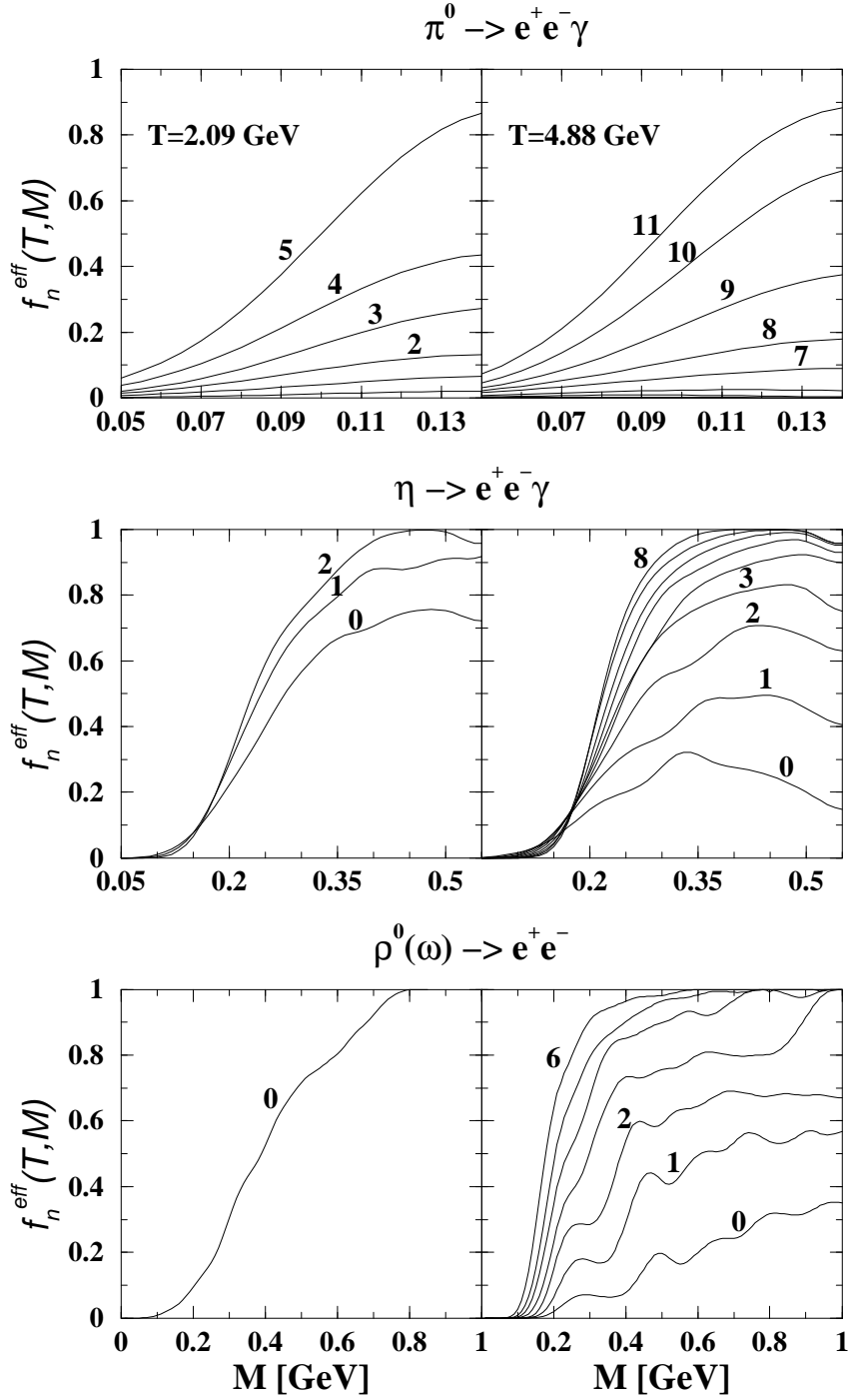}
\end{center}
\caption{The effective acceptance of the DLS detector versus the
invariant dilepton mass, $M$, for different numbers of pions, $n$, produced
in association with the pion $\pi ^{0}$, $\eta $-meson, and $\rho
^{0}(\omega )$-mesons at two highest energies $T=2.09$ and $4.88$ GeV. The
numbers over the curves show the numbers $n$ of the pions.
}
\label{fig4}
\end{figure}

In this context one should be aware that the comparison to the experimental
data is strongly influenced by the acceptance of the DLS detector. In
Appendix 2, we discuss the application of the corresponding filter program 
\cite{W} for the calculation of the experimentally measured cross sections.
In Fig.\ref{fig4} the effective detector efficiency, smeared by a Gaussian
distribution with the standard deviation $\sigma =25~$MeV, is shown as a
function of the dilepton mass $M$ for decays $\pi \rightarrow \gamma
e^{+}e^{-},$ $\eta \rightarrow \gamma e^{+}e^{-},$ and $\rho ^{0}(\omega
)\rightarrow e^{+}e^{-}$ at the two highest proton energies $T=2.09~$GeV and 
$T=4.88~$GeV, where the effects of the multiple pion production are most
important. It can be seen that the effective acceptance decreases with
increasing energy for a fixed number of pions in the final state. On the
other hand, when the number of pions increases, the acceptance increases as
well. This can be interpreted to mean that a larger number of the pions
reduces the available phase space for mesons decaying to the dilepton pairs,
and the decays of such mesons can be detected with better efficiency.

The effect is particularly strong for the $\pi \rightarrow \gamma e^{+}e^{-}$
decay at $T=4.88$ GeV. While the average pion multiplicity $n_{\pi }$ is
around $2,$ the effective acceptance is extremely small below $n\approx 6$. 
The acceptance is not reliable when it is much smaller than
unity. In our case this happens at $n\leq 6$. While the statistical 
distribution gives here, as we expect, reasonable
estimates, the calculation of the part of the cross section connected to the
additional production of $n\leq 6$ pions turns out to be unreliable.
From the other side, at $n\geq 6$ the filter is well defined,
but the binomial distribution gives exponentially small
probabilities. The highest part of the pion spectra corresponding to $%
n\geq 6$ also cannot be calculated accurately. So, we consider the difference
between our results for the pion contributions and those of refs.\cite{ernst,BCEM} 
by about a factor of $3$ and those of ref.\cite{BCM} by about a factor of $6$ as a
conservative estimate for uncertainties inherent in the theoretical
calculations for both, the distribution over the pion multiplicities and the
experimental filter acceptance in the region of small invariant masses. 

At lower energies, these uncertainties practically disappear. In Fig.\ref{fig4}, we
show a plot for the $\pi \rightarrow \gamma e^{+}e^{-}$ decay at $T=2.09$
GeV. It can be seen that now already for $n=0$ the effective acceptance is
no more extremely small. For heavier mesons, the calculation of the
acceptance is safer, which is again connected with less energy available for
the produced mesons and, respectively, a better efficiency for the detection
of the dileptons. 

For the application of the filter we assumed an isotropic
distribution of the particles in the final states in the c.m. frame. This is justified at
small energies $T$. With increasing kinetic energy, the distribution
acquires a bias towards the beam direction. This is an additional source of
uncertainties in the calculations of the filter, which can be important
at energy $T=4.88$ GeV for the pion Dalitz decays.

The many-body phase spaces entering into Eq.(4) are known to be very sharp
functions of the arguments. The Breit-Wigner distribution over the dilepton
mass, $M$, gets therefore an enhancement towards small values of $M$. The
greater the number of pions in the final state, the more important is this
effect. We found that rare processes with probabilities $w_{n}\leq 0.03$
corresponding to large numbers of pions produced in association with the
vector mesons, that should in principle give small contributions to the
total cross sections, become very important at masses $M\leq 200\div 300$
MeV. The spectral functions of the vector mesons are not known well far away
from the vector meson peaks. This effect is thus beyond the scope of the
present model. It should be analysed separately. In the present
calculations, we apply a $3\%$ criterion to the multiple pion
processes: The values $w_{n}$ are set equal to zero every time when $%
w_{n}<0.03.$ This works for the inclusive vector meson production at $%
T=4.88~$GeV. At smaller energies pions are not produced in association with
vector mesons.

\section{Conclusion}

We have considered the dilepton production in $pp$ collisions at BEVALAC
energies. The subthreshold production of vector mesons through the nucleon
resonances is described within the extended VMD model which allows to bring
the transition form factors in agreement with the quark counting rules and
provides an unified
description of the photo- and electroproduction data, $\gamma(\gamma ^{*})N
\rightarrow N^{*},$ the vector meson decays, $N^{*}\rightarrow
N\rho(\omega) $, and the dilepton decays, $N^{*}\rightarrow N\ell ^{+}\ell
^{-}$. The dilepton decay rates are described relativistically using 
kinematically complete phenomenological expressions and numerical results
of ref. \cite{FFKM}. In this context, we discussed also the problem
of double counting and proposed its possible solution. 

The resulting dilepton spectra are reasonably well described at proton energies 
ranging from $T=1.27\div 1.85$GeV. At $T=1.04$ GeV, there exists an overestimation of the
dilepton yield, at $T=2.09$ GeV, we see an underestimation in the vicinity
of the $\omega$-meson peak. At $T=4.88~$GeV we observe an underestimation in
the region of dilepton masses below the $\rho $-peak ($M\approx 400\div 700~$%
MeV). We hope that future experimental investigations will
clarify these problems.

\section{Acknowledgments}
The authors are grateful to L.A. Kondratyuk and A. Matschavariani for useful
discussions of the field-theoretic aspects on the vector meson production, 
H. Matis for providing the DLS filter code and valuable comments,
and U. Mosel for correspondence on the Breit-Wigner formula. Two of us
(M.I.K. and B.V.M.) are indebted to the Institute for Theoretical Physics of
University of Tuebingen for kind hospitality. The work was supported by GSI
(Darmstadt) under the contract T\"{U}F\"{A}ST, by the Plesler Foundation,
and by the Deutsche
Forschungsgemeinschaft under the contract No.~436RUS113/367/0(R). 

\appendix

\section{Breit-Wigner description of resonances with energy-dependent widths}

Let us consider a process with a resonance in the intermediate state. It can
be produced either in a two-body collision or as a result of the decay of a
particle or another resonance. Let the resonance further decay to some
specific channel $i$. The amplitude for the total process, ${\cal M}^{i}$,
i.e. the amplitude for resonance production, propagation, and subsequent
decay to the channel $i$ is a product of the amplitude of its production $%
{\cal M}_{p}$, the resonance propagator, and the amplitude of the resonance
decay ${\cal M}_{d}^{i}$: 
\begin{equation}
{\cal M}^{i}={\cal M}_{p}\frac{1}{p^{2}-m^{2}+\Sigma (p^{2})}{\cal M}%
_{d}^{i}  \label{1n}
\end{equation}
where $p$ is the momentum of the resonance, $m$ is its pole mass, $\Sigma
(p^{2})$ is the resonance self energy. The pole mass $m$ is defined such
that $Re\Sigma (m^{2})=0$. In general, $Re\Sigma (p^{2})$ starts with terms
of the order $O((p^{2}-m^{2})^{2}).$ These terms are further neglected. The
imaginary part of $\Sigma (p^{2})$ is equal to 
\begin{equation}
Im\Sigma (p^{2})=\frac{1}{2}\sum_{i}|{\cal M}_{d}^{i}|^{2}\Phi
_{d}^{i}(p^{2})~,  \label{2n}
\end{equation}
where $\Phi _{d}^{i}(p^{2})$ is the phase space for the resonance decay into
a channel $i$. We can therefore write either 
\begin{equation}
Im\Sigma (p^{2})=\sqrt{p^{2}}\left( \frac{1}{2\sqrt{p^{2}}}\sum_{i}|{\cal M}%
_{d}^{i}|^{2}\Phi _{d}^{i}(p^{2})\right) \equiv \sqrt{p^{2}}\Gamma
_{tot}^{R}(p^{2})  \label{3n}
\end{equation}
or 
\begin{equation}
Im\Sigma (p^{2})=m\left( \frac{1}{2m}\sum_{i}|{\cal M}_{d}^{i}|^{2}\Phi
_{d}^{i}(p^{2})\right) \equiv m\tilde{\Gamma}_{tot}^{R}(p^{2}).  \label{4n}
\end{equation}
Both definitions of the total width, $\Gamma _{tot}^{R}(p^{2})\;$and$\;%
\tilde{\Gamma}_{tot}^{R}(p^{2}),$ can be used in the relativistic
Breit-Wigner formula, but the width should be multiplied by the proper
resonance masses, $\sqrt{p^{2}}$ (running mass) and $m$ (pole mass),
respectively.
\footnote{ In ref.\cite{BCEM} the 
combination $m_{\rho }\Gamma _{e^{+}e^{-}}^{\rho
}(p^{2})$ (physical $\rho $-meson mass and $\Gamma $ without '' $\widetilde{}
$ '') has been substituted into the Breit-Wigner formula. Such a combination
leads to an additional factor $m_{\rho }/M$ in the dilepton production cross
section and, consequently, to an overestimation of the dilepton yield below
the $\rho $-meson peak. We have brought the attention of the authors of ref.%
\cite{BCEM} to this circumstance. Recently, a new paper appeared, ref.\cite
{BCM}, where that inconsistency has been removed.}
The square of the amplitude $M^{i}$, integrated over the phase space of the
final particles with momenta, $p_{p}^{k}$ and $p_{d}^{l},$ and normalized by
the corresponding factors for the initial particles, gives either a cross
section (two initial particles) or a width (one initial particle).

For the scattering problem, the cross section has the form 
\begin{eqnarray}
d\sigma &=&\frac{1}{j2E_{1}2E_{2}}|{\cal M}^{i}|^{2}(2\pi )^{4}\delta
^{(4)}(p_{1}+p_{2}-\Sigma _{k}p_{p}^{k}-\Sigma _{l}p_{d}^{l}) \nonumber \\
&\times & \prod_{k}\frac{%
d^{3}p_{p}^{k}}{(2\pi )^{3}2E_{p}^{k}}\prod_{l}\frac{d^{3}p_{d}^{l}}{(2\pi
)^{3}2E_{d}^{l}},  \label{5n}
\end{eqnarray}
where $j$ is the flux of the incoming particles.

Further, we introduce two $\delta $-functions corresponding to momentum
conservation in the processes of production and decay of the resonance and
the running mass $M$ of the intermediate resonance 
\begin{equation}
\begin{array}{c}
\delta ^{(4)}(p_{1}+p_{2}-\Sigma _{k}p_{p}^{k}-\Sigma _{l}p_{d}^{l})= \\ 
=\int \delta ^{(4)}(p_{1}+p_{2}-\Sigma _{k}p_{p}^{k}-p)\delta
^{(4)}(p-\Sigma _{l}p_{d}^{l})d^{4}p\int \delta (M^{2}-p^{2})dM^{2}
\end{array}
\end{equation}
and obtain 
\begin{equation}
\sigma =\int \sigma (M^{2})\frac{1}{\pi }\frac{M\Gamma _{i}(M)dM^{2}}{%
(M^{2}-m^{2})^{2}+(M\Gamma _{tot}^{R}(M))^{2}}
\end{equation}
where 
\begin{equation}
M\Gamma _{i}(M)\equiv m\tilde{\Gamma}_{i}(M)\equiv \frac{1}{2}|{\cal M}%
_{d}^{i}|^{2}\Phi _{d}^{i}(M^{2}).
\end{equation}
Here, one should also use the partial widths $\Gamma _{i}(M),$ $\tilde{\Gamma%
}_{i}(M)$ with the proper masses $M,$ $m.$ The similar arguments apply in
case of the decay problem.\medskip

\section{Effective filter function}

A comparison to the DLS data requires to take the experimental detector
efficiency into account. For this purpose a filter function is provided by
the DLS collaboration. In particular at large $T$ (e.g. $T=4.88~$GeV) this
filter function is not a small correction to the theoretical calculations
but is crucial for the comparison to data. Thus, in this appendix we discuss
the influence of the detector filter in our analysis.

In terms of the c.m. frame momentum variables, the filter function can be
rewritten as 
\begin{equation}
f(p_{T},y,M)=f(p_{T}^{*},y^{*}+y_{c},M)
\end{equation}
where $p_{T}^{*}=p_{T}$ is the transverse momentum of the dilepton pair, $%
y_{c}$ is the rapidity of the c.m. frame $L^{*}$ with respect to the
laboratory frame $L$ of the colliding nucleons, 
\begin{equation}
y_{c}=\frac{1}{2}ln(\frac{\sqrt{s}+\sqrt{s-4m_{N}^{2}}}{\sqrt{s}-\sqrt{%
s-4m_{N}^{2}}}),
\end{equation}
$T$ is the proton kinetic energy in the $L$ frame. The distribution of
dileptons in the c.m. frame $L^{*}$ is isotropic. This is a universal
feature which does not depend on the specific type of the reactions and is
connected to the form of the cross section (4) only.
So, we work with the filter function averaged over the angles
in the $L^{*}$ frame: 
\begin{equation}
f(p^{*},y_{c},M)=\int_{-1}^{+1}\frac{dcos\vartheta }{2}%
f(p_{T}^{*},y^{*}+y_{c},M)
\end{equation}
where 
\begin{eqnarray*}
p_{T}^{*} &=&p^{*}sin\vartheta , \\
y^{*} &=&\frac{1}{2}ln(\frac{\epsilon ^{*}+p_{||}^{*}}{\epsilon
^{*}-p_{||}^{*}}),
\end{eqnarray*}
and 
\begin{eqnarray*}
p_{||}^{*} &=&p^{*}cos\vartheta , \\
\epsilon ^{*} &=&\sqrt{M^{2}+p^{*2}}.
\end{eqnarray*}
The problem reduces to finding the dilepton distribution over the dilepton
momentum $p^{*}$ in the c.m. frame $L^{*}$.

The probability distribution of the dilepton momentum in the $L^{*}$ frame
for the direct decays $V\rightarrow e^{+}e^{-}$ is given by 
\begin{equation}
dW(p^{*})=\sum_{n=0}^{N_{\pi }}w_{n}D_{n}\Phi _{2}(\sqrt{s}%
,M,M_{X})dM_{X}^{2}\Phi _{2+n}(M_{X}...)
\end{equation}
where 
\begin{eqnarray*}
\Phi _{2+n}(M_{X}...) &=&\Phi _{2+n}(M_{X},m_{N},m_{N},\mu _{\pi },...,\mu
_{\pi }), \\
D_{n} &=&\Phi _{3+n}^{-1}(\sqrt{s},m_{N},m_{N},M,\mu _{\pi },...,\mu _{\pi
}).
\end{eqnarray*}
The effective filter function can be calculated as follows 
\begin{equation}
f^{eff}(T,M)=\sum_{n=0}^{N_{\pi }}w_{n}f_{n}^{eff}(T,M)=\int
dW(p^{*})f(p^{*},y_{c},M).  \label{DIRECT}
\end{equation}
The value of $M_{X}$ is integrated out within the limits $2m_{N}+n\mu _{\pi
}\leq M_{X}\leq $ $\sqrt{s}-M.$ The momentum of the dilepton pair $%
p^{*}=p^{*}(\sqrt{s},M,M_{X})$ is given by 
\[
p^{*}(\sqrt{s},M,M_{X})=\frac{\sqrt{(s-(M+M_{X})^{2})(s-(M-M_{X})^{2})}}{2%
\sqrt{s}}. 
\]
A $100\%\ $detector efficiency would yield $f(p_{T},y,M)=1$ and $%
f^{eff}(T,M)=1$ in virtue of Eq.(\ref{sw=1}). The values $f_{n}^{eff}(T,M)$
are plotted in Fig.\ref{fig4} for the $\rho ^{0}(\omega )\rightarrow e^{+}e^{-}$
decays for different values of $n$ at energies $T=2.09$ and $4.88$ GeV.

For the Dalitz decays ${\cal M}\rightarrow {\cal M}^{\prime }e^{+}e^{-}$,
the probability distribution for the dilepton energy $\epsilon ^{*}$ in the
c.m. frame $L^{*}$ can be written as follows 
\begin{eqnarray}
dW(\epsilon ^{*})&=&\sum_{n=0}^{N_{\pi }}w_{n}D_{n}\int_{(2m_{N}+n\mu _{\pi
})^{2}}^{(\sqrt{s}-M)^{2}}dM_{X}^{2}\int_{-1}^{+1}\frac{dcos\chi }{2}\delta
(\epsilon ^{*}-\eta k)d\epsilon ^{*} \nonumber \\
&\times &\Phi _{2}(\sqrt{s},\mu _{{\cal M}%
},M_{X})\Phi _{2+n}(M_{X}...).  \label{XXXX}
\end{eqnarray}
Here, $k$ is the four momentum of the dilepton pair, $\mu _{{\cal M}}$ is
the mass of the meson ${\cal M}$ in the reaction $pp\rightarrow {\cal M}X\ $%
with the subsequent decay ${\cal M}\rightarrow {\cal M}^{\prime }e^{+}e^{-}.$
In the $L^{*}$ frame, the vector $\eta $ is defined as $\eta =(1,{\bf 0}).$
The coefficients $D_{n}$ are defined as before.

Now we should pass to the rest frame $L^{**}$ of the meson ${\cal M\ },$
where $\eta =\gamma (1,-{\bf n}v).$ The $\gamma $-factor and the velocity $v$
of the meson ${\cal M}$ in the $L^{*}$ frame are determined by equations 
\begin{eqnarray*}
\gamma \mu _{{\cal M}} &=&(s+\mu _{{\cal M}}^{2}-M_{X}^{2})/(2\sqrt{s}), \\
v\gamma \mu _{{\cal M}} &=&p^{*}(\sqrt{s},\mu _{{\cal M}},M_{X}).
\end{eqnarray*}
The unit vector ${\bf n}$ shows in the direction of the meson velocity in
the $L^{*}$ frame. In the meson rest frame, $L^{**}$, the dilepton pair has
momentum $k=(\epsilon ^{**},{\bf n}^{\prime }p^{**})$ where 
\begin{eqnarray*}
\epsilon ^{**} &=&(\mu _{{\cal M}}^{2}+M^{2}-\mu _{{\cal M}^{\prime
}}^{2})/(2\mu _{{\cal M}}), \\
p^{**} &=&p^{*}(\mu _{{\cal M}},\mu _{{\cal M}^{\prime }},M).
\end{eqnarray*}
The function $p^{*}(...)$ is defined earlier. The unit vector ${\bf n}%
^{\prime }$ shows in the direction of the dilepton pair momentum in the $%
L^{**}$ frame. In Eq.(\ref{XXXX}) the value $\chi $ is the angle between the
directions of the meson velocity in the $L^{*}$ frame and velocity of the
dilepton pair in the $L^{**}$ frame, so that $cos\chi ={\bf nn}^{\prime }$
and therefore $\eta k=\gamma (\epsilon ^{**},vp^{**}cos\chi )$.

In Eq.(40) the integral over the angle $\chi $ is evaluated
explicitly, and we obtain 
\begin{equation}
\frac{dW(\epsilon ^{*})}{d\epsilon ^{*}}=\sum_{n=0}^{N_{\pi }}w_{n}D_{n}%
\frac{\pi \mu _{{\cal M}}}{2\sqrt{s}p^{**}}\int_{(2m_{N}+n\mu _{\pi
})^{2}}^{(\sqrt{s}-M)^{2}}dM_{X}^{2}\theta (\epsilon ^{*},M_{X})\Phi
_{2+n}(M_{X}...).
\end{equation}
where

\begin{equation}
\theta (\epsilon ^{*},M_{X})=\left\{ 
\begin{array}{ll}
1, & \gamma (\epsilon ^{**}-vp^{**})\leq \epsilon ^{*}\leq \gamma (\epsilon
^{**}+vp^{**}), \\ 
0, & {\normalsize otherwise.}
\end{array}
\right.
\end{equation}
The effective filter function can now be calculated to be 
\begin{equation}
f^{eff}(T,M)=\sum_{n=0}^{N_{\pi }}w_{n}f_{n}^{eff}(T,M)=\int dW(\epsilon
^{*})f(p^{*},y_{c},M).  \label{DALITZ}
\end{equation}

The values $f_{n}^{eff}(T,M)$ are plotted in Fig.\ref{fig4} for the $\pi ^{0}(\eta
)\rightarrow e^{+}e^{-}\gamma $ decays for different values of $n$ at
energies $T=2.09$ and $4.88$ GeV.

It is now sufficient to multiply the differential cross section (\ref{V1})
with the corresponding effective filter function $f^{eff}(T,M)<1$ in order
to compare the calculations with the experiment. For the evaluation of the
direct contributions one should use expression (\ref{DIRECT}), while for the
Dalitz decays one should use expression (\ref{DALITZ}). The function $%
f^{eff}(T,M)$ is given by a two-dimensional integral in Eq.(\ref{DIRECT})
and by a three-dimensional integral in Eq.(\ref{DALITZ}).

\end{document}